# Measurement of DC Magneto-Optical Kerr Effect with Sensitivity of $10^{-7}$ Rad/$\sqrt{\text{Hz}}$

Junying Ma[1], Feng Gu[1], Ying Xu[1], Jiaming Le[1], Fanlong Zeng[1], Yizheng Wu[1], Chuanshan Tian[1,*]

[1]*State Key Laboratory of Surface Physics and Department of Physics, Fudan University, Shanghai 200433, China*

*Electronic mail: cstian@fudan.edu.cn

**Abstract:**

A high-sensitive DC magneto-optical Kerr effect (MOKE) apparatus is described in this letter. Via detailed analysis on several dominating noise sources, we've proposed solutions that significantly lower the MOKE noise, and a sensitivity of $1.5 \times 10^{-7}$ rad/$\sqrt{\text{Hz}}$ is achieved with long-term stability. The sensitivity of the apparatus is tested by measuring a wedge-shaped Ni thin film on $SiO_2$ with Ni thickness varying from 0 to 3 nm. A noise floor of $1.5 \times 10^{-8}$ rad is demonstrated. The possibility of further improving sensitivity to $10^{-9}$ rad via applying ac modulation is also discussed.

## I. INTRODUCTION

Polarization measurement has gained broad applications in many research topics, examples include magnetic anisotropy,[1] spin dynamics[2] in magnetic material, birefringence in chiral media[3] and electro-optic sampling technique.[4] In magnetism, one convenient and popular analytical tool is based on the magneto-optical effect, which alters the polarization of the reflected (Kerr effect) and the transmitted light (Faraday effect) through the asymmetric dielectric tensor induced by the magnetization.[1] Since its first application to surface magnetism,[5] magneto-optical Kerr effect (MOKE) has been developed as a non-intrusive and versatile probe for remote measurements on static or dynamic properties of spin systems with very high sensitivity, e.g. spin Hall effect,[6,7] ultrafast spin dynamics,[8] imaging magnetic domain and nanostructure,[9,10] as well as magneto-optic information storage.[11] However, because the polarization of light is very sensitive to a large variety of noise sources, it's difficult to achieve a sensitivity of $10^{-7}\,\mathrm{rad}/\sqrt{\mathrm{Hz}}$ in MOKE measurement, especially in DC detection scheme.[1,7,12,13,14] This hampers the application of MOKE in many newly emerging subjects, such as spin Hall effect,[6] time-reversal symmetry breaking states in superconductor,[15,16] where a sensitivity of $10^{-7}\sim10^{-8}$ rad is urgently needed.

In polarization measurement, a significant difficulty in pushing the AC/DC MOKE sensitivity to $10^{-7}\ \mathrm{rad}/\sqrt{\mathrm{Hz}}$ is the overwhelming noise from reciprocal effects including linear birefringence and thermal fluctuations.[16] In order to suppress the noise from the reciprocal effects, zero loop-area Sagnac interferometry, which measures the time-reversal-symmetry-breaking (TRSB) Kerr effect, has been employed to promote the MOKE detection limit. A sensitivity as small as $10^{-7}\mathrm{rad}/\sqrt{\mathrm{Hz}}$ has been achieved for polar MOKE through phase modulation of the probe light at ~5 MHz.[16,17] Although it can be well adapted to the measurement of polar magnetization, high-sensitive probe of the in-plane magnetization, i.e. longitudinal- and transverse-MOKE, remains challenging. By inserting reflections optics to fold the oblique-incident beam path backward, the modified Sagnac interferometer can be applied to measure the in-plane magnetization with a sensitivity of $10^{-6}$ rad/$\sqrt{\mathrm{Hz}}$.[18,19] Another widely used approach is based on high-frequency modulation of the sample magnetization, which may reach the lowest noise floor of several $10^{-9}$ rad with sensitivity of ~$10^{-7}$rad/$\sqrt{\mathrm{Hz}}$.[6,20,21] Unfortunately, to date, the detection limit of DC MOKE is limited to $\sim 2\times 10^{-7}$ rad (0.01mdeg),[14] which hinders the study of the static in-plane and out-of-plane magnetic properties, such as TRSB states in $Sr_2RuO_4$[16] and $PrOs_4Sb_{12}$[22]. More importantly, a MOKE apparatus with state-of-the-art DC detection capability can set a thorough grounding for further improvement of MOKE sensitivity when AC modulation scheme is implanted. Therefore, breaking the bottleneck in polarization measurement is urgently needed.

In this letter, we report a general solution for achieving a DC MOKE sensitivity of $1.5\times 10^{-7}$ rad/$\sqrt{\mathrm{Hz}}$ with long-time stability using the balanced detection scheme. Three noise sources were identified dominating the MOKE signal-to-noise ratio, namely, drift of laser cavity modes, temperature-induced strain in polarizing optics, and turbulence of airflow, which cause the polarization fluctuations in the optical measurement. After stabilizing these variables, the apparatus was used to measure the hysteresis loop of a wedge-like Ni film with thickness varying from 0 to 3 nm. An RMS noise of $1.5\times 10^{-8}$ rad was demonstrated with an averaging time of 200s each point. Although not yet implemented in this study, further improvement of sensitivity is feasible via AC modulation with lock-in detection.

## II. EXPERIMENT AND RESLUTS

The experimental setup is sketched in Fig. 1(a). The longitudinal MOKE geometry was chosen for demonstration. (The scheme is also valid for polar- and transverse-MOKE by varying the direction of external magnetic field.) The light source was a commercial HeNe laser (12 mW, R-30993, Newport, $\lambda = 632.8\mathrm{nm}$) with linearly polarized output. The laser beam passed through a zero-order half-wave plate (HWP1) and a Glan-Taylor polarizer P (GT10, Thorlabs) with polarization aligned perpendicular to the optical plane ($s$-polarization). To improve the extinction ratio, a piece of sapphire window is placed after the polarizer such that the laser beam is reflected from the window surface at Brewster angle. A $p$-polarized component appears associated with the dominant $s$-polarized component after reflection from a magnetic sample due to the magneto-optical Kerr effect, where the ratio of their electric fields $E_p/E_s$ equals to the Kerr rotation angle $\theta_k$. The polarization change was measured by a balanced detection setup consisting of a zero-order half-wave plate (HWP2), a Wollaston prism (WP10, Thorlabs) and a balanced detector (Nirvana Model 2007, New Focus).[6]

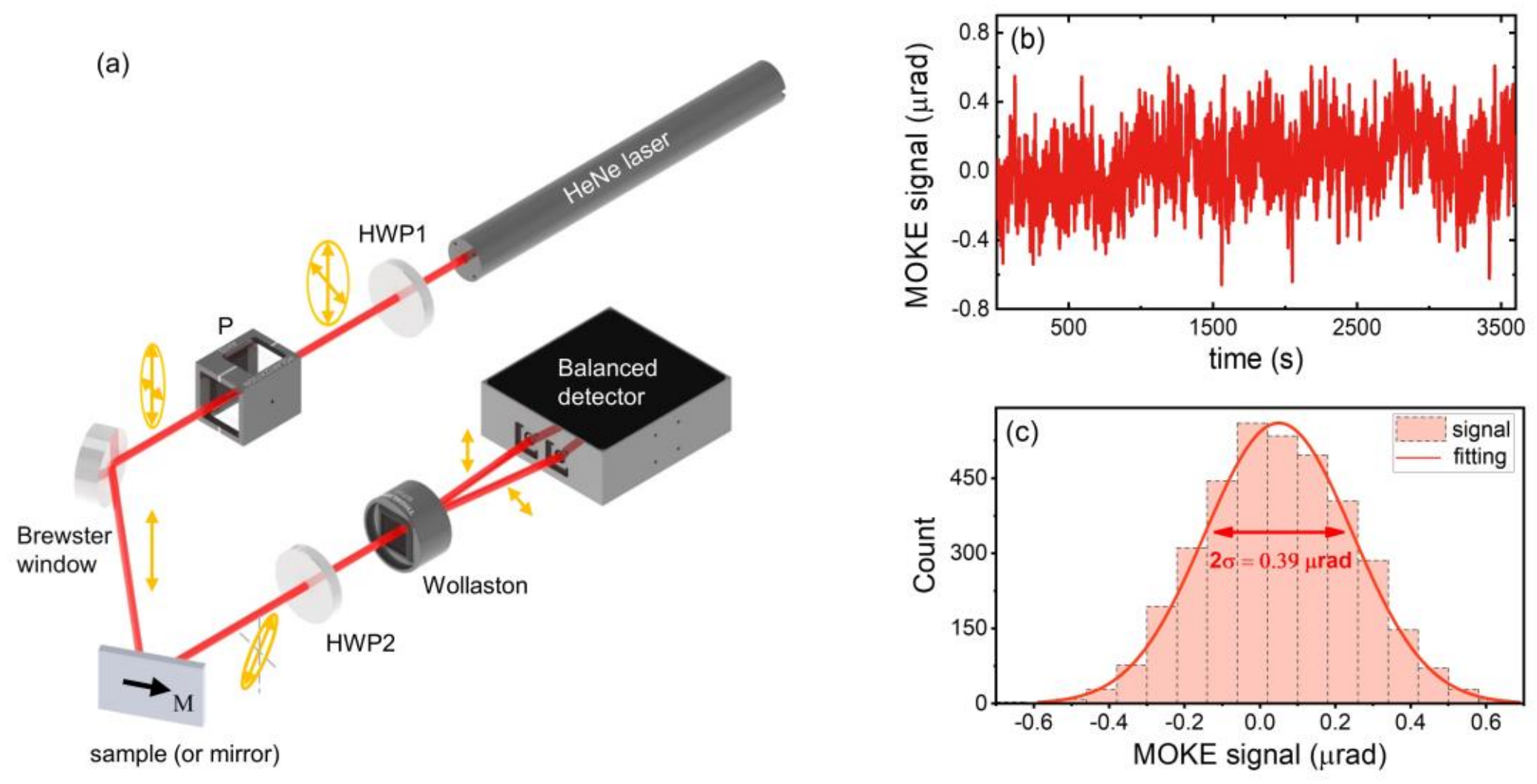

**Fig. 1.** (a) Sketch of the DC-MOKE setup. HWP1 and HWP2 stand for half-wave plates, and P is the polarizer. The arrows illustrate the polarization state after each optics. (b) Fluctuation of MOKE signal in an hour after control of the temperature within $\pm 1$ mK for the laser tube and polarizing optics in a sealed environment. (c) Statistic analysis of the MOKE noise in (b) using Gaussian distribution.

To suppress the polarization noise in the setup, three dominating factors were identified and properly taken care of, i.e. the temperature-induced variation of the laser cavity length and birefringence in the polarizing optics, and the airflow turbulence. The temperature fluctuation of the laser and the polarizing optics were controlled within ±1 mK using a home-built temperature controller. To avoid air turbulence, all of the optical components except for the laser were placed in a closed black box, with the entrance aperture of the laser beam sealed with an optical window. As a result, the equivalent noise of 6.3μV/$\sqrt{\text{Hz}}$ was achieved for the output voltage from the balanced detector over an hour, as shown in Fig. 1(b), which corresponds to a MOKE measurement sensitivity (RMS) of 1.5× $10^{-7}$ rad/$\sqrt{\text{Hz}}$. Given the short-term and long-time stability, a measurement of Kerr rotation as small as $1.5 \times 10^{-8}$ rad and $2.5 \times 10^{-9}$ rad can be realized for integrating time of 100 seconds and 1 hour, respectively. In the following, we will discuss in detail how different noise sources affect MOKE sensitivity.

**III. NOISE ANALYSIS**

Considering $s$-polarized light being reflected from a magnetic sample, the resultant $s$- and $p$-polarized components are rotated by an angle of $\alpha$ ($\alpha$~45°) using the half-wave plate (HWP2 in Fig. 1(a)), and then interfere constructively and destructively in the two detection arms after the Wollaston prism, respectively. The intensity difference between the two arms is given by[23]

$$\Delta I \approx (-\cos 2\alpha + 2\theta_k \sin 2\alpha) \times I_0 \quad (1)$$

Here, $I_s = |E_s|^2$ is the intensity of the reflected $s$-polarized light. Via fine-tuning of the angle $\alpha \to 45°$, the first term on the right-hand side may vanish, and we have $\theta_k = \Delta I / 2I_s$. Note that the common mode fluctuation from the laser intensity is canceled in $\Delta I$. Still, the polarization noise of the light persists, contributing to the fluctuation of the MOKE signal ($\Delta\theta_k$).

To show how thermal fluctuations affect polarization measurement, we modulate the temperature of laser and polarizing optics and record the MOKE signal concurrently. Figure 2(a) shows the MOKE signal fluctuating along with the laser intensity, as the laser temperature is drifting. The seemingly correlation actually does not mean the intensity fluctuation is the noise source, because the variation $\Delta\theta_k/\theta_k$ (~33%) is much larger than the intensity noise $\Delta I_s/I_s$ (~0.26%). Furthermore, the amplitude of the MOKE fluctuation remains the same regardless of fine-tuning of the balance between the two split beams, namely tuning the value of $\alpha$, suggesting the intensity noise again is not the cause (more detailed discussion can be found in the supplementary material).

The fluctuations of the MOKE signal and the laser intensity in Fig. 2(a) are actually both the consequence of the variation of the laser cavity modes. It is well known that the adjacent longitudinal modes, labeled as $s$-mode and $p$-mode in Fig. 2(b)), in red (632.8 nm) He-Ne lasers are orthogonally polarized.[24] To demonstrate the change of the two modes versus cavity length, we chose another He-Ne laser without polarization control in the cavity. As shown in Fig. 2(b), the output energy alternates between the two polarizing modes with precise synchronization. In other words, the power changes of the two polarization states are out of phase.[25] Despite polarizing optics is generally placed inside the laser, the unwanted weak $p$-modes remain in the cavity even though the net gain factor is much smaller. As the laser tube temperature is drifting, its cavity length ($L$) varies, and the frequency-modes sweep across the Neon gain curve[26]. During the mode-sweeping process, change of the dominating $s$-modes and the residual $p$-modes satisfy the relation of $\Delta E_p/\bar{E}_p = -\Delta E_s/\bar{E}_s$. It is then readily to derive from Eq. (1) the corresponding variation of the MOKE signal with details given in supplementary material:

$$\Delta\theta_k \approx \frac{1}{\sqrt{\beta}} \times \left(\frac{\Delta E_s}{\bar{E}_s} - \frac{\Delta E_p}{\bar{E}_p}\right) = \frac{2}{\sqrt{\beta}} \times \frac{\Delta E_s}{\bar{E}_s} \tag{2}$$

Here, $\Delta E_i$ and $\bar{E}_i$ represents the variation and average of the electric field ($E_i$), respectively, and $\beta \sim 1 \times 10^5$ is the extinction ratio of the polarizer before sample (P in Fig. 1(a)) without the Brewster window. Given the intensity variation of 0.26% in the mode-sweeping process, we find $\Delta E_s/\bar{E}_s$ = 0.13%. Using Eq. (2), one may readily estimate the change of the MOKE signal is $8.2 \times 10^{-6}$ rad. It agrees nicely with the experimental result of $8 \times 10^{-6}$ rad shown in Fig. 2(a).

According to Eq. (2), it is clear that to reduce the MOKE noise caused by the laser, one needs to avoid the mode-sweeping process via stabilization of the cavity length and to improve the extinction ratio ($\beta$). As shown in Fig. 2(c), the laser intensity fluctuation is reduced down to 0.02% when the temperature fluctuation of the laser tube is kept within $\pm 1$ mK. Meanwhile, the sapphire Brewster window inserted after the polarizer P increases the extinction ratio via attenuating the unwanted $p$-polarized component in the reflected beam. Figure 2(d) compares MOKE noise with and without the Brewster window, where the temperature of the laser is stabilized, yet which of the polarizing optics is not controlled. Obviously, the polarization noise has been largely suppressed by the Brewster window.

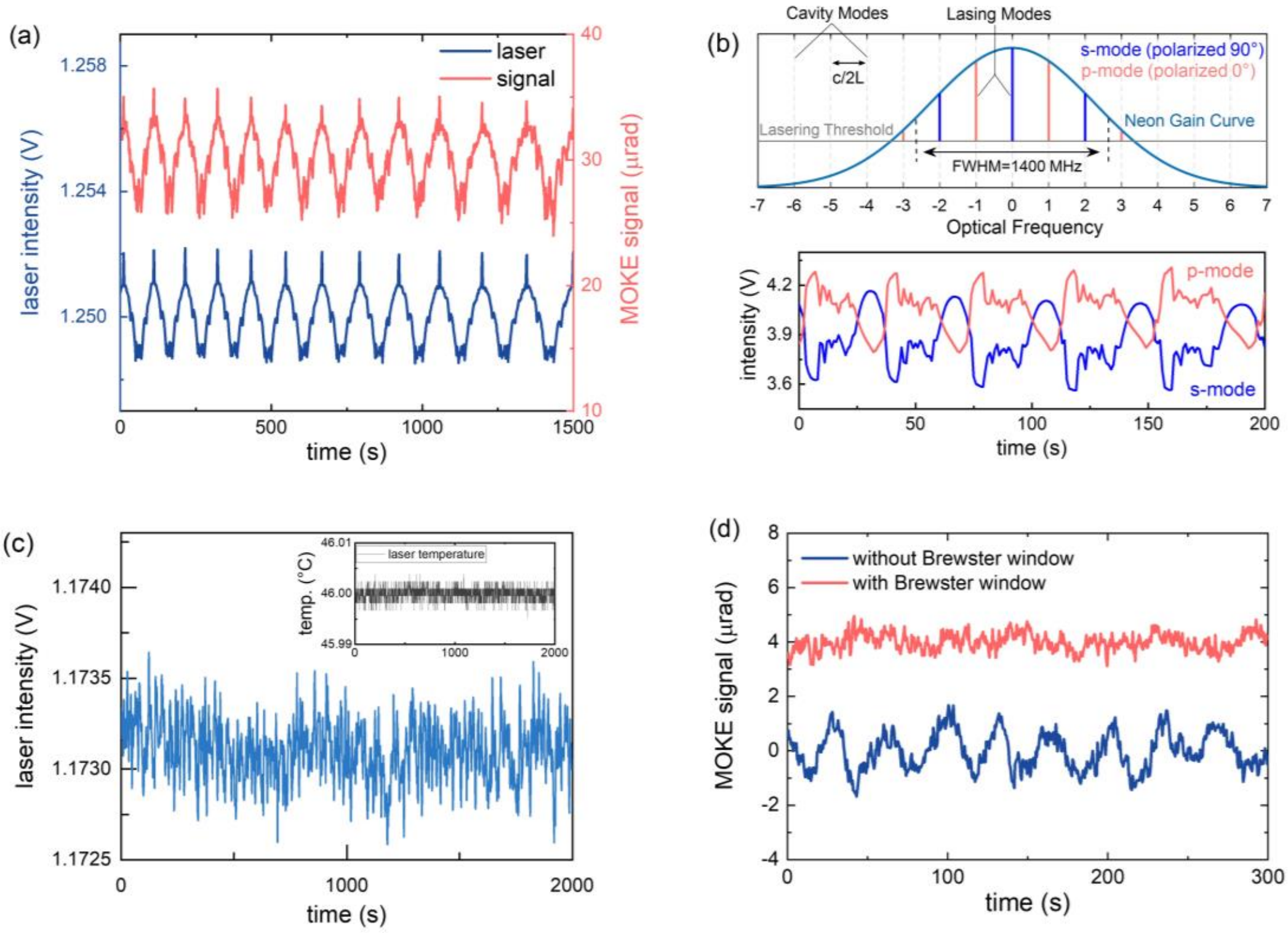

**FIG. 2.** (a) MOKE signal (red line) fluctuates along with the laser intensity (blue line) as the laser tube temperature is drifting. The fine spectral feature is the fingerprints of the gain medium. (b) Top, mode structure of a red (632.8 nm) He-Ne laser. The adjacent longitudinal modes, labeled as *s*-mode (blue line) and *p*-mode (red line), are orthogonally polarized. Bottom shows the measured intensity variances of s-mode (blue) and p-mode (red) are out of phase in a HeNe laser with cavity length of 25 cm. (c) The fluctuation of laser intensity after temperature control of $\pm 1$ mK for the laser tube (inset). (d) Comparison of the polarization noise with (red) and without (blue) the Brewster window.

It is important to point out that, besides the laser fluctuation, the temperature-induced birefringence and the air turbulence also contribute notably to the polarization noise. The former mainly affects the long-term stability, while the latter induces the high-frequency noise. To evaluate the impact of temperature fluctuation on the polarizing optics, we intentionally oscillate the temperature of the polarizer and the Wollaston prism slowly while recording MOKE signal. The results are depicted in Fig. 3(a) and 3(b), which shows that a temperature variation of $\pm 0.05$ K on the Glan-Taylor polarizer and the Wollaston prism causes approximately $\pm 2 \times 10^{-6}$ rad change in the Kerr signal, suggesting the necessity of stabilizing the temperature within a few mK to achieve long-term sensitivity better than $10^{-7}$ rad. On the other hand, airflow disturbance is another primary noise source as it influences both the polarization and pointing of the laser beam. Fig. 3(c) compares the noise level in a sealed box and with the top cover open. In the latter case, the noise increases by a factor of 5 in an open environment. Also, in an open environment, the continuously-varying and inhomogeneous air temperature may induce birefregence in optics that gives rise to instability of polarization. Thus, to achieve high-accurate MOKE measurement, one needs to control the temperature stability down to a few mK and contain the optical path in a closed environment.

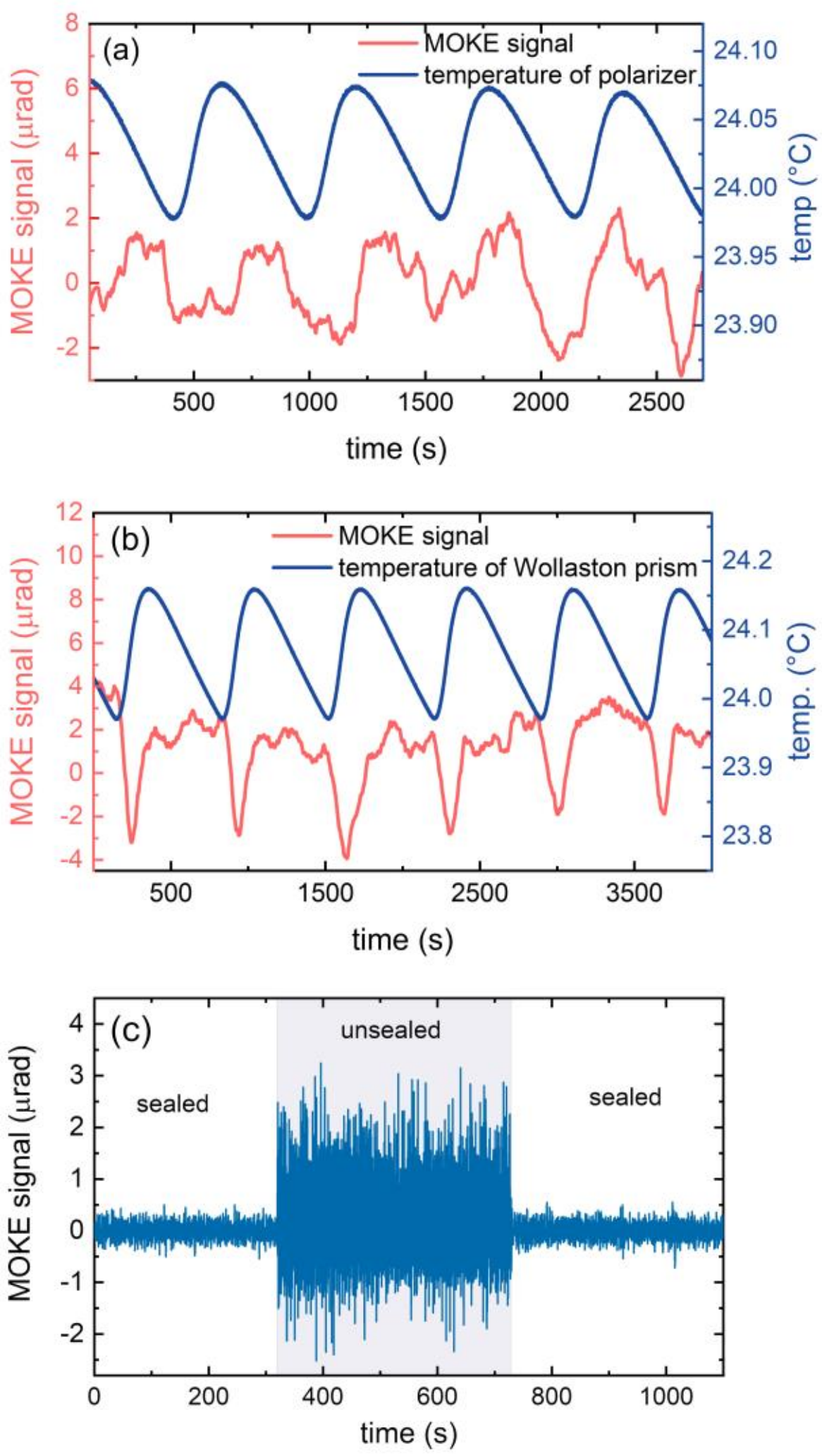

**FIG. 3.** (a) and (b) The variation of DC-MOKE signal (red line) when modulating the temperature (blue

line) of polarizer (a) and Wollaston (b), respectively. (c) Comparison of MOKE noise in sealed and unsealed condition after subtracting the drifting background.

## IV. HYSTERESIS LOOPS OF A WEDGE-SHAPED NI THIN FILM

After careful control of the noise sources mentioned above, the sensitivity of the apparatus is tested by measuring a wedge-shaped Ni thin film on $SiO_2$ substrate with Ni thickness varying from 0 to 3 nm. The magnetic hysteresis loops are shown in Fig. 4(a) measured at five positions on the sample with different thickness of Ni. The data for the bare substrate and those for Ni thickness at 3 nm and 2.2 nm were recorded with averaging time of 200 s/point, while the loops of 2.8 nm- and 2.4 nm-thick Ni were taken using 0.5 s integrating time per point. To characterize the noise level, we show in Fig. 4(b) the hysteresis loop of the bare $SiO_2$ substrate. The RMS noise of the loop reaches $1.5 \times 10^{-8}$ rad.

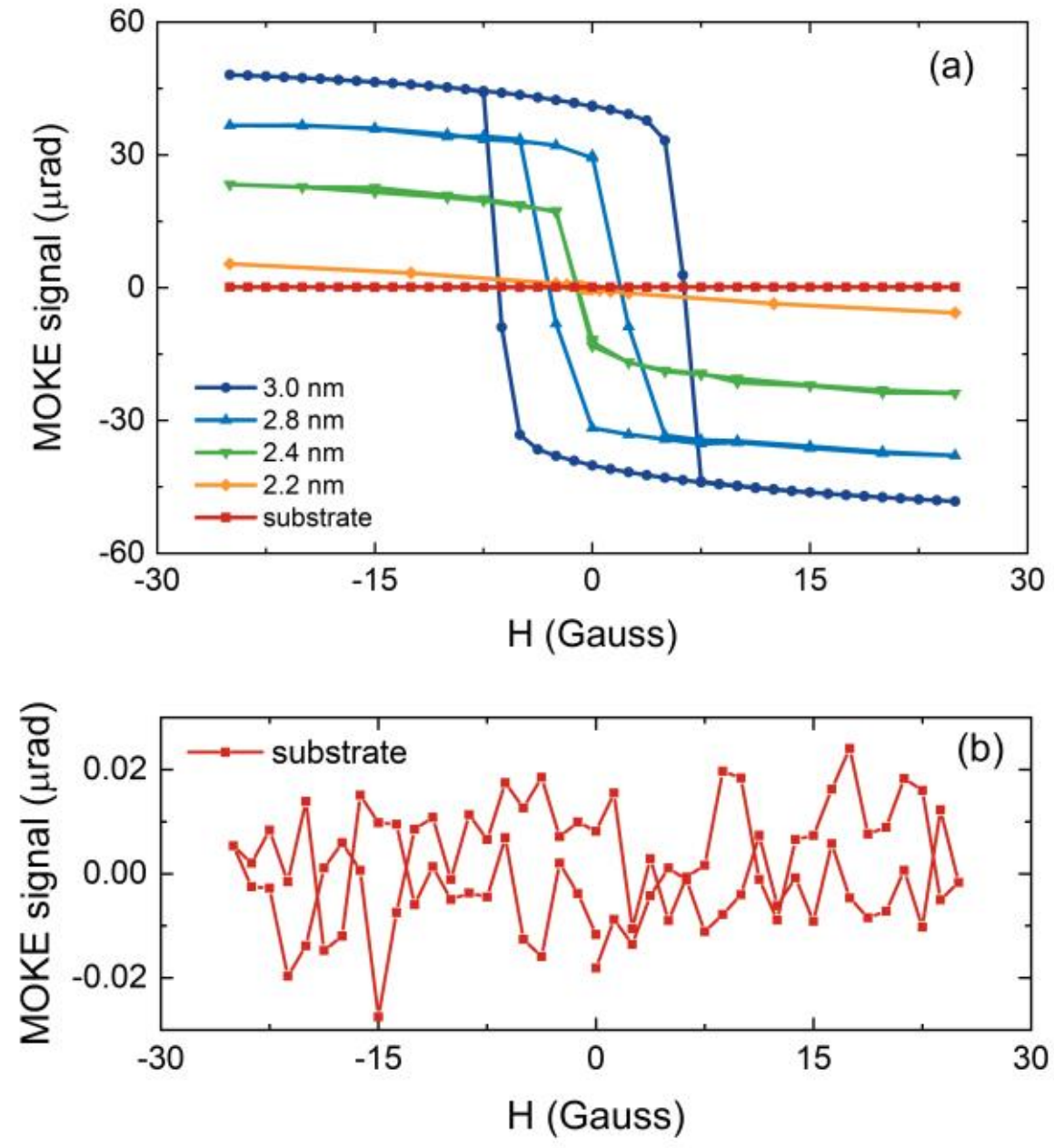

**FIG. 4.** (a) Hysteresis loops at five different positions of a wedge-shaped Ni thin film on $SiO_2$ substrate. (b) Noise measured at the bare $SiO_2$ substrate.

## V. DISCUSSION ON AC MODULATION SCHEME

With the DC polarization noise reduced down to $1.5\times 10^{-7}$ rad$/\sqrt{\text{Hz}}$, the MOKE sensitivity may be further improved by AC modulation associated with lock-in technique.[6] We then record in Fig. 5 the noise spectrum of the apparatus between 200 Hz and 3 kHz. In the region of 2.1-3 kHz, the noise floor decreases to $0.3\mu\mathrm{V}/\sqrt{\mathrm{Hz}}$, which corresponds to $7\times 10^{-9}\,\mathrm{rad}/\sqrt{\mathrm{Hz}}$. It is 20 times better than the DC case. Therefore, a polarization sensitivity of $7\times 10^{-9}$ rad is achievable with 1s integrating time if the probe beam or the sample is modulated at frequency above 2.1 kHz. Benefiting from the long-term stability, a few-nano rad sensitivity is possible via increasing the integrating time.

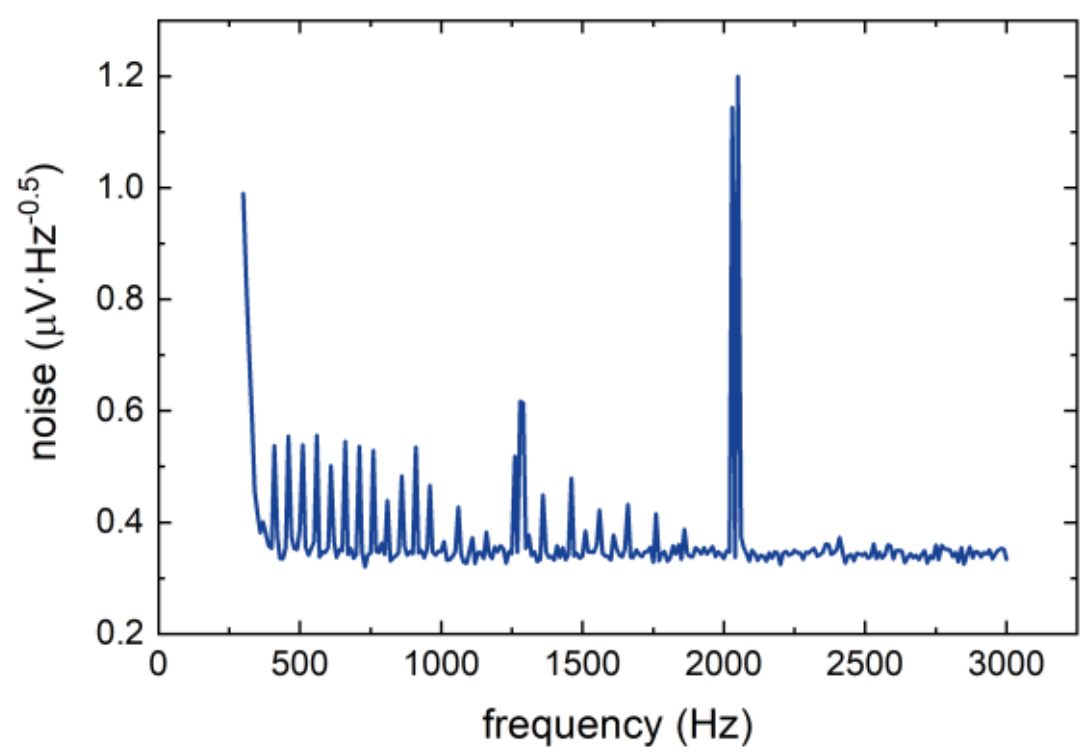

**FIG. 5.** The noise spectrum of our MOKE apparatus measured by SR830 lock-in amplifier.

## VI. CONCLUSION

In conclusion, we have demonstrated a long-term stable DC MOKE apparatus with sensitivity of $1.5\times 10^{-7}\,\mathrm{rad}/\sqrt{\mathrm{Hz}}$. We analyzed three noise sources in the polarization measurement including dirft of laser cavity mode, temperature-induced birefregence and airflow. Through high-accuracy temperature control of the laser cavity and those polarizing optics in a sealed condition, polarization noise has been greatly suppressed. As a result, a MOKE signal from Ni thin-film as small as $1.5\times 10^{-8}$ rad can be resolved in the DC measurement scheme. Our work provides a general solution for precision measurement of light polarization not only for TRSB spin states in magnetic and novel quantum materials, but also for polarization-sensitive physics in a wide range of research topics.

## SUPPLEMENTARY MATERIAL

See supplementary material for the complete noise analysis on HeNe laser.

## ACKNOWLEDGEMENT

C.T. acknowledges the funding support from the National Natural Science Foundation of China (No. 12125403 and No. 11874123) and the Shanghai Science and Technology Committee (No. 20ZR1406000).

## AUTHOR DECLARATIONS

Conflict of Interest

The authors have no conflicts to disclose.

## DATA AVAILABILITY

The data that support the findings of this study are available from the corresponding author upon reasonable request.